\documentclass[11pt]{article}
\usepackage{authblk}
\usepackage[margin=0.8in]{geometry}
\usepackage{amsmath,amssymb,amsthm}
\usepackage{graphicx} % Required for inserting images
\usepackage{caption}
\usepackage{subcaption}
\usepackage{hyperref}
\usepackage{cleveref}
\usepackage{braket}
\usepackage{comment}
\usepackage{xcolor}
\usepackage{cases}
\usepackage{algorithm}
\usepackage{algpseudocode}
\usepackage{booktabs}
\usepackage{cite}

\newcommand{\E}{\mathbb{E}}

\newtheorem{theorem}{Theorem}
\newtheorem{lemma}{Lemma}

\title{Quantum Box-Muller Transform}
\author[1]{Dinh-Long Vu\thanks{long-vu@nus.edu.sg}}
\author[2,1]{Hitomi Mori\thanks{hmori.academic@gmail.com}}
\author[1]{Patrick Rebentrost\thanks{cqtfpr@nus.edu.sg}}
\affil[1]{Centre for Quantum Technologies, National University of Singapore, Singapore, Singapore}
\affil[2]{Graduate School of Engineering Science, Osaka University, Osaka, Japan}

\date{}

\begin{document}
\maketitle
\abstract{
The Box-Muller transform is a widely used method to generate Gaussian samples from uniform samples. Quantum amplitude encoding methods encode the multi-variate normal distribution in the amplitudes of a quantum state. This work presents the Quantum Box-Muller transform which creates a superposition of binary-encoded grid points representing the multi-variate normal distribution. The gate complexity of our method  depends on quantum arithmetic operations and, using a specific set of known implementations, the complexity is quadratic in the number of qubits. 
We apply our method to Monte-Carlo integration, in particular to the estimation of the expectation value of a function of Gaussian random variables.
Our method implies that the state preparation circuit used multiple times in amplitude estimation requires only quantum arithmetic circuits for the grid points and the function, in addition to a single controlled rotation.
We show how to provide the expectation value estimate with an error that is exponentially small in the number of qubits, similar to the amplitude-encoding setting with error-free encoding. 
}

%\abstract{
%The Box-Muller transform is a widely used method to generate Gaussian samples from uniform distributions. This work presents a quantum counterpart of this technique. Conventional state preparation methods encode the multi-variate normal distribution in the amplitudes of the state vector, also called analog encoding \cite{Mitarai_2019}. The Quantum Box-Muller transform creates a binary encoding of a grid of points representing the multi-variate normal distribution using quantum arithmetic circuits.  The gate complexity of our algorithm depends on how these arithmetic operations are implemented. For a specific set of implementations adopted in this paper, the complexity is quadratic in the number of qubits. 
 %quadratic in the number of qubits. Moreover, within the context of quantum Monte-Carlo integration, we demonstrate how our method is used to estimate the expectation value of a function of Gaussian random variables with error that is exponentially small in the number of qubits. As a result, quantum estimation of such random variables becomes state-preparation free and quantum-arithmetic only.
%}

\section{Introduction}
Normal distributions are the building block of many stochastic process and mathematical finance models, from the basic model of Black and Scholes~\cite{Black1973,Merton1973} to stochastic volatility models such as the Heston model and the SABR model. Being able to efficiently access normal distributions is thus indispensable for potential quantum applications in finance and other domains. But first let us ask: what exactly do we mean by accessing a normal distribution? The natural answer \cite{grover2002creating,rattew2022preparing} is to prepare a pure quantum state $|\psi\rangle$ such that the measurement probabilities in the computational basis form a discretized normal distribution, i.e.,
\begin{align}
    |\psi\rangle = \frac{1}{\mathcal{N}}\sum_{j=0}^{N-1} \sqrt{f(x_j)} |j\rangle,\label{eq_first}
\end{align} 
where $f$ denotes the probability density function (pdf) of the normal distribution, $x_j$ is the real-number outcome represented by the binary string $j$, $\mathcal{N}=\sqrt{\sum_{j}f(x_j)}$ is a normalization factor, and $N=2^n$ with the number of qubits $n$.
This preparation has also been investigated approximately with variational methods \cite{Zoufal_2019}.
An alternative method based on arithmetic circuits is to
compute the pdf at every point $j$. These circuits allow us to prepare
\begin{align}
    |\psi^{pdf}\rangle = \frac{1}{\sqrt{N}}\sum_{j=0
    }^{N-1}  |j\rangle \left | \sqrt{f(x_j)}\right \rangle, \label{eq_simplealternative}
\end{align}
where $ \sqrt{f(x_j)}$ is stored in a fixed-point representation in a register of qubits.
This state can be seen as a precursor to the state in Eq.~(\ref{eq_first}), and a controlled rotation and amplification can be used to obtain the state in Eq.~(\ref{eq_first}).
A subsequent question arises: Can the quantum task consider a set of numbers $Z^{(j)}$ that represent the normal distribution by a grid of points transformed from the uniform points $x_j$? Such a method encodes the normal distribution via the state
\begin{align}
    |\psi^*\rangle = \frac{1}{\sqrt{N}}\sum_{j=0
    }^{N-1}  \left |Z^{(j)} \right \rangle, \label{second}
\end{align}
where $Z^{(j)}$ are given in fixed-point binary representation. If the ultimate goal of our quantum algorithm is to estimate the expectation value of some function of a normally distributed variable, then \textit{a priori} there is no reason to favor one over the other. Interestingly enough, the first direction in Eq.~\eqref{eq_first} has so far been taken as the standard for the Gaussian state preparation task. In this work, we present, to our knowledge, the first work in the direction of Eq.~\eqref{second}, utilizing the Box-Muller transform to construct points corresponding to \textit{two}-dimensional normal distributions. 

Our tool is the Box-Muller transform, a common method for generating normal random variables in classical computing. Starting with two samples $U$ and $V$ from independent uniform random variables  on $[0,1]$, it can be shown that $(Z_1,Z_2)$  correspond to samples from a standard two-dimensional Gaussian distribution if
\begin{align}
    Z_1 &= \sqrt{-2\ln U}\sin (2\pi V)\;,\\
    Z_2 &=  \sqrt{-2\ln U}\cos (2\pi V)\;.
\end{align}
The translation to quantum is straightforward: the uniform variables can be prepared using Hadamard gates while the arithmetic operations for the cosine, multiplication, natural logarithm, and square root all have their quantum counterparts. The resulting quantum Box-Muller Monte-Carlo estimation is free from state-preparation subroutines for Gaussian amplitudes and relies solely on arithmetic operations, controlled rotations, and amplitude estimation. One may be concerned about the large cost incurred by these quantum arithmetic operations, and we consider the method to be indeed not suited for near-term devices. We argue that other parts of the algorithm (such as computation of the payoff function) often also require arithmetic operations, those operations can be compiled and optimized together with the quantum Box-Muller part. In addition, the sample points generated from the Box-Muller transform have so far not been discussed in quantum computing and may find application beyond Monte Carlo estimation. 

We further compare to existing work and highlight the benefits of our work. Among methods that rely on quantum arithmetic operations, our method is  economic with only a constant number of operations (i.e., ln, sin, and so on). The Grover-Rudolph method \cite{grover2002creating} encodes the probability distribution by an iterative process, where a qubit is added in each iteration and increases the resolution of the distribution. In each step, access to an operation is required that computes  subintegrals of $f$. Similarly, the Kitaev-Webb method is based on the Grover-Rudolph method, but is designed specifically to implement a normal distribution \cite{kitaev2009}. As these methods require integration using quantum arithmetic circuits, the gate complexity becomes substantially high. 
%The Grover-Rudolph method in particular, requires a number of integrations that is exponentially large in the number of qubits. 
Adiabatic state preparation \cite{rattew2022preparing} starts from the uniform superposition state and adiabatically evolves towards the target distribution state. To implement the adiabatic time evolution, the access to the oracle $O_f$ is assumed. 
%This oracle uses arithmetic operations for computing the Gaussian density function $f$ and the number of queries to the oracle is $O(\delta^{-2})$ where $\delta$ is the deviation from an ideal unitary in terms of a spectral distance.  
In the black-box method \cite{Grover_2000}, the target function is encoded in qubits first using the oracle $O_f$ and then exported to the amplitude. 
%This process requires $O(\sqrt{n})$ arithmetic operations $\arcsin$ however, with $n$ being the number of qubits.
These methods involve the iterations and thus require repeated access to the oracle, while our approach avoids this repetition.

%Secondly, even though there are several Gaussian state preparation methods without quantum arithmetic operations \cite{Chakrabarti_2021,Rattew2021efficient,9259933,Iaconis_2024,Moosa_2023,rosenkranz2024quantum,mcardle2022quantumstatepreparationcoherent,mori2024efficientstatepreparationmultivariate}, such circuit is specifically tailored for the standard normal distribution. To obtain a generic normal distribution, one has to apply a linear transformation on the components of the state vector, which inevitably involves multiplication and addition. It turns out that the arithmetic operations required by the Box-Muller transform are not much more costly to implement than the multiplication itself, see \cref{section:arithmetic}. There are other methods, such as the quantum Generative Adversarial Network (qGAN) \cite{Zoufal_2019}, which can simulate a complex distribution. However, the classical cost of training the neural network could be very high. 

Unlike the methods introduced so far, there also exist methods that avoid arithmetic operations. In Fourier series loader (FSL) \cite{Moosa_2023}, the coefficients of the Fourier series approximation of the target function are loaded first, and the Fourier series is obtained through the quantum Fourier transform.
Similarly, another method based on linear combination of unitary (LCU) \cite{rosenkranz2024quantum} has recently been proposed as a multivariate function state preparation method that implements Fourier or Chebyshev series. This method uses block-encodings of each term of Fourier or Chebyshev series and combines them using LCU.
Instead of arithmetic operations, both the FSL and LCU-based methods require computation of coefficients of Fourier/ Chebyshev series approximation and loading of these coefficients on a quantum circuit, which can be costly especially when extended to multivariate case.
Lastly, the work \cite{mcardle2022quantumstatepreparationcoherent} utilizes an easily-implementable block-encoding and performs Quantum Signal Processing (QSP) on the ancilla qubit to construct the target function, followed by exact amplitude amplification to obtain the target state.
This method is also arithmetic-free, but requires the computation of angle parameters and the extension to the multi-variable case is not straightforward \cite{mori2024efficientstatepreparationmultivariate}. It is also possible to implement the multi-variate normal distribution using univariate state preparation $D$ times, but in this case arithmetic operation is inevitable to incorporate the correlation using Cholesky decomposition. 

Last but not least, the quantum Monte-Carlo integration consists of three parts: state preparation, payoff calculation, and amplitude estimation. If one tries to avoid quantum arithmetics by choosing amplitude-encode the distribution in the first step, there are still the arithmetic computations of the second step, i.e., the computation of the random variable or the payoff function. There are several methods to compute the payoff function without relying on arithmetic operations, however, they come with tradeoffs and restrictions. The approximation trick in \cite{Woerner_2019,Stamatopoulos_2020} applies to piece-wise linear payoff functions but it decreases the quadratic speed-up to $O(\epsilon^{-\frac{2d+2}{2d+3}})$ where $d$ is the degree of the approximating polynomial.  In \cite{Herbert_2022}, the idea is to decompose the payoff function into Fourier modes and to apply amplitude estimation to each mode. This decomposition only works given a restricted smoothness condition on the function, namely it must be continuous in value and first derivative, and has second and third derivatives that are piece-wise continuous and bounded. Such a requirement is not satisfied by many payoff functions of interest in finance.  

The manuscript is organized as follows. In \cref{section:bm} we present the classical Box-Muller and its quantum counterpart. In \cref{section:algo} we analyze the approximation error of the quantum Box-Muller transform. In \cref{section:arithmetic} we quantify the cost of the arithmetic operations involved. Finally, \cref{sec:conclusion} concludes our work.

%See also the method of \cite{Stamatopoulos2024derivativepricing,stamatopoulos2024quantumriskanalysisfinancial}, which uses quantum signal processing~\cite{low_methodology_2016,gilyen_quantum_2019} to build approximating polynomials.

\section{Box-Muller transform}\label{section:bm}
\subsection{Classical Box-Muller method}
To review the classical Box-Muller transform, recall the change of variables rule. If $X$ is a random variable with probability density function (PDF) $f_X$ and $Y=g(X)$ is another random variable, then the PDF of $Y$ is given by
\begin{align}
    f_Y(y) = f_X(g^{-1}(y))\left|\frac{d g^{-1}(y)}{dy}\right|.
\end{align}
Now let $Z = (Z_1,Z_2)$ be a standard two-dimensional normal distribution with PDF
\begin{align}
    f_Z(z_1,z_2) = \frac{1}{2\pi}\exp
    \left(-\frac{1}{2}(z_1^2+z_2^2)\right).
\end{align}
Denote by $(R,\Theta)$ the polar coordinates i.e. $Z_1,Z_2= (R\sin \Theta,R\cos\Theta)$. Then the joint PDF factorizes into PDFs of $R$ and $\Theta$
\begin{align}
    f_{R,\Theta}(r,\theta)=\frac{1}{2\pi}r\exp(-r^2/2) :=f_\Theta (\theta) f_R(r)\;.
\end{align}
That is, $R$ and $\Theta$ are independently distributed. Moreover if we denote $U := \exp(-R^2/2)$ and $V := \Theta/2\pi$  then it is clear that $U$ and $V$ are uniform distributions in $[0,1]$. Correlated random variables $X_1$ and $X_2$ that follow
\begin{align}
    f_X(x_1,x_2)=\frac{1}{2\pi\sqrt{1-\rho^2}}\exp\left(-\frac{1}{2(1-\rho^2)}(x_1^2-2\rho x_1x_2+x_2^2)\right),
\end{align}
where $\rho$ is the correlation coefficient, can also be generated by defining 
\begin{align}
        X_1=Z_1, \quad X_2=\rho Z_1+\sqrt{1-\rho^2}Z_2.
    \end{align}
\begin{theorem}[Classical Box-Muller transform] Suppose $U$ and $V$ are independent uniform random variables in $[0,1]$. Let
\begin{align}
    Z_1 &= \sqrt{-2\ln U}\sin (2\pi V)\;,\label{Z1}\\
    Z_2 &=  \sqrt{-2\ln U}\cos (2\pi V)\;.\label{Z2}
\end{align}
    Then $Z_1$ and $Z_2$ are independent standard normal random variables.
    Moreover, let
    \begin{align}
        X_1=Z_1,\quad X_2=\rho Z_1+\sqrt{1-\rho^2}Z_2.
    \end{align}
    Then $X_1$ and $X_2$ are correlated standard normal random variables with the correlation coefficient $\rho$.
\end{theorem}
With $a:= \sin(2 \pi v)$ and $b:= \cos(2 \pi v)$, note that the Jacobian is
\begin{align}
    \begin{vmatrix}
        \frac{\partial x_1}{\partial u}&\frac{\partial x_1}{\partial v}\\
        \frac{\partial x_2}{\partial u}&\frac{\partial x_2}{\partial v}
    \end{vmatrix}&=
    \begin{vmatrix}
        \frac{-a}{u\sqrt{-2\ln u}} &2\pi b \sqrt{-2\ln u}\\
        \frac{-(\rho a+\sqrt{1-\rho^2}b)}{u\sqrt{-2\ln u}}&2\pi\sqrt{-2\ln u}(\rho b-\sqrt{1-\rho^2}a )
    \end{vmatrix}\nonumber=\frac{2\pi\sqrt{1-\rho^2}}{u}.
\end{align}
The probability density function is
\begin{align}
    \tilde{f}_X(u,v)&=f_X\left(\sqrt{-2\ln u}a,\sqrt{-2\ln u}(\rho a+\sqrt{1-\rho^2}b )\right)\nonumber=\frac{u}{2\pi\sqrt{1-\rho^2}}
\end{align}
In the change of probabilistic variables, these two factors cancel out, resulting in
\begin{align}
    f_X(x_1,x_2)dx_1dx_2=\tilde{f}_X(u,v)\begin{vmatrix}
        \frac{\partial x_1}{\partial u}&\frac{\partial x_1}{\partial v}\\
        \frac{\partial x_2}{\partial u}&\frac{\partial x_2}{\partial v}
    \end{vmatrix}dudv=dudv.\label{eq:change of variables}
\end{align}
This simple property will be useful in later discussions.

\subsection{Quantum Box-Muller method}
In this section, we extend the classical Box-Muller method to quantum settings. Similar to the classical Box-Muller method, we start from a quantum state composed of two uniform superpositions:
\begin{align}
    \frac{1}{N}\sum_{j,k=0}^{N-1}\ket{j}\ket{k}.
\end{align}
Then we add ancilla qubits to encode two variables that follow the uncorrelated bivariate normal distribution. The target state is given by
\begin{align}
    \frac{1}{N}\sum_{j,k=0}^{N-1}\ket{j}\ket{k}\ket{Z_1^{(j,k)}}\ket{Z_2^{(j,k)}},
\end{align}
where 
\begin{align}
    &Z_1^{(j,k)}=\sqrt{-2\ln U^{(j)}}\sin(2\pi V^{(k)}),\\
    &Z_2^{(j,k)}=\sqrt{-2\ln U^{(j)}}\cos(2\pi V^{(k)}).
\end{align}
Preparing this state can be considered as a quantum version of Box-Muller method.

\begin{figure}[t]
    \centering
    \includegraphics[width=0.9\linewidth]{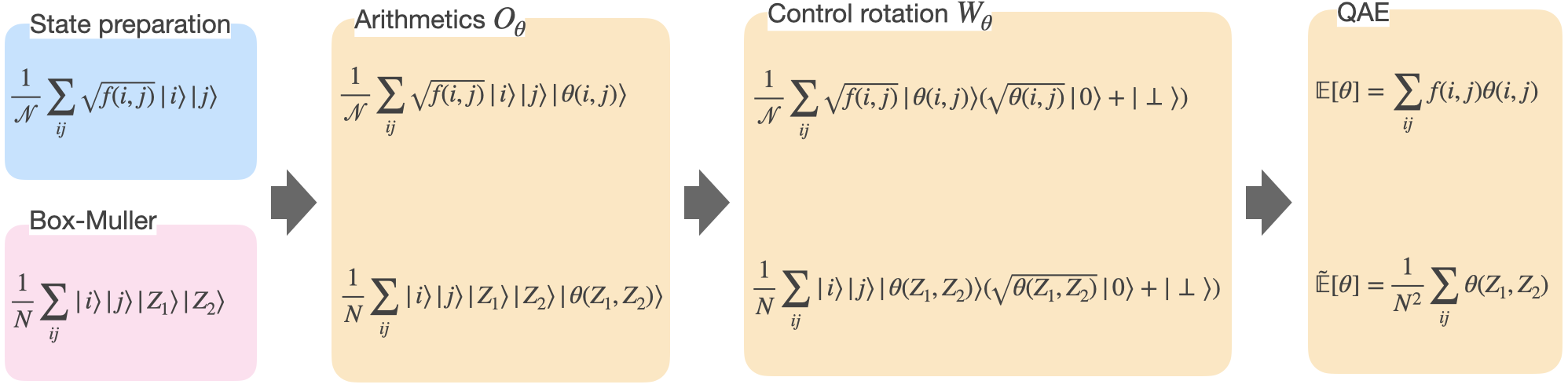}
    \caption{Comparison of the standard method (state preparation) and our method (Box-Muller). Here, the random variable $\theta$ is given via $O_\theta:\ket{i}\ket{j}\ket{0}\to\ket{i}\ket{j}\ket{\theta(i,j)}$ and the controlled rotation is given via the operation $W_\theta:\ket{\theta}\ket{0}\to\ket{\theta}(\sqrt{\theta}\ket{0}+\sqrt{1-\theta}\ket{1})$.}
    \label{fig:methods}
\end{figure}

Suppose we want to obtain $\mathbb{E}[\theta(Z_1,Z_2)]$ as a result of Monte Carlo simulation.
Here we assume $\theta\in[0,1]$ for illustration purpose (this case corresponds to Algorithm \ref{alg:case1}).
We add another register of ancilla qubits and store $\theta(Z_1,Z_2)$ as
\begin{align}
    \frac{1}{N}\sum_{j,k=0}^{N-1}\ket{j}\ket{k}\ket{Z_1^{(j,k)}}\ket{Z_2^{(j,k)}}\ket{\theta(Z_1^{(j,k)},Z_2^{(j,k)})}.
\end{align}
Then we add another register and apply a controlled rotation to export the function $\theta$ in the ancilla as
\begin{align}
    \frac{1}{N}\sum_{j,k=0}^{N-1}\ket{j}\ket{k}\ket{Z_1^{(j,k)}}\ket{Z_2^{(j,k)}}\ket{\theta(Z_1^{(j,k)},Z_2^{(j,k)})}\left(\sqrt{\theta(Z_1^{(j,k)},Z_2^{(j,k)})}\ket{0}+\sqrt{1-\theta(Z_1^{(j,k)},Z_2^{(j,k)})}\ket{1}\right).
\end{align}
The probability of post-selecting the ancilla state $\ket{0}$ is given by
\begin{align}
    \frac{1}{N^2}\sum_{j,k=1}^{N-1}\theta(Z_1^{(j,k)},Z_2^{(j,k)}),\label{eq:expectation}
\end{align}
which is the estimate for $\mathbb{E}[\theta(Z_1,Z_2)]$.
Amplitude estimation allows the efficient estimation of this post-selection probability.
Figure \ref{fig:methods} illustrates the essence of our method compared to the standard state preparation approach.
Note that in the classical algorithm, the samples are obtained from the uniform distribution, while the quantum algorithm repeatedly prepares and processes a uniform superposition.

Using our framework, one can also easily implement correlated standard normal variables by encoding the following $X_1$ and $X_2$ with the correlation coefficient $\rho\in[0,1]$:
\begin{align}
    \frac{1}{N}\sum_{j,k=0}^{N-1}\ket{j}\ket{k}\ket{X_1^{(j,k)}}\ket{X_2^{(j,k)}},
\end{align}
where 
\begin{align}
    X_1=Z_1,\quad X_2=\rho Z_1+\sqrt{1-\rho^2}Z_2.\label{eq:linear_transformation}
\end{align}
In the following sections, we focus on the bivariate case, but the generalization to the $D$-variable case is straightforward. We prepare $\lceil D/2\rceil$ quantum Box-Muller transforms to obtain a standard $D$-dimensional normal distribution. A distribution of a given mean and covariance matrix can then be obtained from the standard distribution by a linear transformation.

\section{Algorithms}\label{section:algo}
We consider three types of assumption on $\theta$ in parallel with Ref. \cite{Montanaro_2015} as summarized in Table \ref{tab:asm}.
\begin{table}[ht]
\centering
\begin{tabular}{@{\extracolsep\fill}lccc}
\toprule%
Case & Assumption & Algorithm  & Theorem\\
\midrule
1 & $\theta\in[0,1]$,  & Algorithm \ref{alg:case1}& Theorem \ref{thm:case1}  \\
2 & $\theta\geq 0$, $\E[\theta^2]\le B^2$ & Algorithm \ref{alg:case2}& Theorem \ref{thm:case2}\\
3 & $\mathrm{Var}[\theta]\le \sigma^2$ & Algorithm \ref{alg:case3}& Theorem \ref{thm:case3}\\
\bottomrule
\end{tabular}
\caption{Summary of assumptions. Despite the apparent similarity with table 1 of \cite{Montanaro_2015}, our theorems analyze the error with respect to the true expectation value. That is, we take into account both the discretization and truncation error that are inherent in the finite encoding of a continuous random variable.}\label{tab:asm}
\end{table}
\subsection{Preliminary}
We recite here some standard results from \cite{Montanaro_2015} that will later be used in our main theorems.
\begin{lemma}\label{lemma:case1}
Let $\mathcal{A}$ be a quantum algorithm acting on $n$ qubits with a measurement on the last $k$ qubits. Denote by $\theta(x)\in [0,1]$ the output value of $\mathcal{A}$ when the measurement outcome $x\in \lbrace 0,1\rbrace^k$ is received. Let $W$ be the unitary operator acting on $k+1$ qubits defined by 
\begin{align}
    W|x\rangle |0\rangle = |x\rangle (\sqrt{\theta(x)}|0\rangle +\sqrt{1-\theta(x)}|1\rangle).
\end{align}
There exists an algorithm that uses $O(t\log (1/\delta))$ copies of $\mathcal{A}$ and $W$, and outputs an estimate $\hat{\mu}$ such that 
\begin{align}
    |\hat{\mu}-\E[\theta]|\leq C\left(\frac{\sqrt{\E[\theta]}}{t}+\frac{1}{t^2}\right)
\end{align}
with probability at least $1-\delta$, where $C$ is a universal constant.
\end{lemma}
\begin{lemma}\label{lemma:case2} Assume that $\theta(x) \geq 0$ and $\E[\theta^2]\leq B^2$, there exists an algorithm that uses $\tilde{O}(\log(1/\delta)/\epsilon)$ copies of $\mathcal{A}$ and $W$, and outputs an estimate $\hat{\mu}$ such that
\begin{align}
    |\hat{\mu}-\E[\theta]|\leq \epsilon(B+1)^2
\end{align}
with probability at least $1-\delta$.
\end{lemma}
\subsection{Case 1: $\theta\in[0,1]$}
The algorithm for the case where the range of $\theta$ is bounded within $[0,1]$ is given in Algorithm \ref{alg:case1}. Here, we assume independent random variables $Z_1,Z_2\sim\text{Norm}(0,1)$ for simplicity of the discussion.
Note that the linear transformation $(z_1,z_2)\to (x_1,x_2)$ in \cref{eq:linear_transformation} only alters the constant prefactor and does not affect the following discussion.
\begin{algorithm}[h]
\caption{Estimate $\mu=\E[\theta(Z_1,Z_2)]$ for $\theta\in[0,1]$}\label{alg:case1}
\begin{algorithmic}[1]
\Require A function $\theta\in[0,1]$ such that $\left|\frac{\partial \theta}{\partial z_1}\right|+\left|\frac{\partial \theta}{\partial z_2}\right|$ and $\left|\frac{\partial^2 \theta}{\partial z_1^2}\right|+\left|\frac{\partial^2 \theta}{\partial z_1\partial z_2}\right|+\left|\frac{\partial^2 \theta}{\partial z_2^2}\right|$ are bounded, truncation points $u_\text{min},u_\text{max}$ such that $U\in[u_\text{min},u_\text{max}]$, $\delta>0$, integers $N,t$.
\begin{enumerate}
\item Using Hadamard gates, prepare the uniform superposition as
\begin{align}
    \frac{1}{N}\sum_{j,k=0}^{N-1}\ket{j}\ket{k}.
\end{align}
\item Add registers and encode $Z_1=\sqrt{-2\ln U^{(j)}}\sin(2\pi V^{(k)}),Z_2=\sqrt{-2\ln U^{(j)}}\cos(2\pi V^{(k)})$ as
\begin{align}
    \frac{1}{N}\sum_{j,k=0}^{N-1}\ket{j}\ket{k}\ket{Z_1^{(j,k)}}\ket{Z_2^{(j,k)}},
\end{align}
where $U^{(j)}=u_\text{min}+\frac{u_\text{max}-u_\text{min}}{N-1}j,V^{(k)}=\frac{k}{N-1}$.
\item Add a register and apply $O_\theta:\ket{Z_1}\ket{Z_2}\ket{0}\to\ket{Z_1}\ket{Z_2}\ket{\theta(Z_1,Z_2)}$ to obtain 
\begin{align}
    \frac{1}{N}\sum_{j,k=0}^{N-1}\ket{j}\ket{k}\ket{Z_1^{(j,k)}}\ket{Z_2^{(j,k)}}\ket{\theta(Z_1^{(j,k)},Z_2^{(j,k)})}.\label{eq:function_query}
\end{align}
\item Add a register and apply control rotation $W_\theta:\ket{\theta}\ket{0}\to\ket{\theta}(\sqrt{\theta}\ket{0}+\sqrt{1-\theta}\ket{1})$ to obtain
\begin{align*}
    \frac{1}{N}\sum_{j,k=0}^{N-1}\ket{j}\ket{k}\ket{Z_1^{(j,k)}}\ket{Z_2^{(j,k)}}\ket{\theta(Z_1^{(j,k)},Z_2^{(j,k)})}\left(\sqrt{\theta(Z_1^{(j,k)},Z_2^{(j,k)})}\ket{0}+\sqrt{1-\theta(Z_1^{(j,k)},Z_2^{(j,k)})}\ket{1}\right).
\end{align*}
\item Apply $O(t\log(1/\delta))$ iterations of QAE to $\ket{0}$ and output $\hat{\mu}$ as the estimate of
\begin{align}
    \frac{1}{N^2}\sum_{j,k=0}^{N-1}\theta(Z_1^{(j,k)},Z_2^{(j,k)}).
\end{align}
\end{enumerate}
\end{algorithmic}
\end{algorithm}
The error bound for Algorithm \ref{alg:case1} is given by the following Theorem \ref{thm:case1}.

\begin{theorem}\label{thm:case1}
Let $\epsilon,C_1>0$,
%u_\text{min}=C_1\epsilon/2,u_\text{max}=\exp(-C_1\pi\epsilon/2)$ 
and assume a function $\theta\in[0,1]$ such that $\left|\frac{\partial \theta}{\partial z_1}\right|+\left|\frac{\partial \theta}{\partial z_2}\right|$ and $\left|\frac{\partial^2 \theta}{\partial z_1^2}\right|+\left|\frac{\partial^2 \theta}{\partial z_1\partial z_2}\right|+\left|\frac{\partial^2 \theta}{\partial z_2^2}\right|$ are bounded by a  constant. There exists an algorithm 
%Then Algorithm \ref{alg:case1} 
that outputs the estimate $\hat{\mu}$ such that
\begin{align}
    |\E[\theta(Z_1,Z_2)]-\hat{\mu}|\le C_1\epsilon+\frac{C_2}{N^2\epsilon^2}+C_3\left(\frac{\sqrt{\mu}}{t}+\frac{1}{t^2}\right),\label{eq:error_algo1}
\end{align}
with probability at least $1-\delta$, for some constants $C_2,C_3>0$. To achieve $|\E[\theta(Z_1,Z_2)]-\hat{\mu}|\le \epsilon$, it suffices to take $N\in O(\epsilon^{-3/2})$, number of qubits $n=\lceil\log_2 N\rceil$ and $t\in O(\epsilon^{-1})$, which corresponds to a number of queries to $O_\theta$ and $W_\theta$ in $O(\epsilon^{-1}\log(1/\delta))$ .
\end{theorem}

\proof
Define $u_\text{min}=C_1\epsilon/2$ and $u_\text{max}=\exp(-C_1\pi\epsilon/2)$. 
Define the integral $I := \mathbb{E}[\theta(Z_1,Z_2)] \equiv L(0,\infty)$ and the truncated integral
\begin{equation}
    I(l,L) := \int\int_{l^2\leq z_1^2+z_2^2\leq L^2}\theta(z_1,z_2)f_Z(z_1,z_2)dz_1dz_2.
\end{equation}
The total error in Eq. \eqref{eq:error_algo1} can be decomposed into three different errors
\begin{align}
    \left|\mathbb{E}[\theta(Z_1,Z_2)]-\hat{\mu}\right|\le\epsilon_\text{trunc}+\epsilon_\text{disc}+\epsilon_\text{QAE},
\end{align}
where $\epsilon_\text{trunc},\epsilon_\text{disc},\epsilon_\text{QAE}$ are defined as follows:
\begin{align}
    \epsilon_\text{trunc} := &\left|I-I(l,L)\right|,\\
    \epsilon_\text{disc} := &\left|I(l,L)-\frac{1}{N^2}\sum_{i,j=1}^{N-1}\theta(z_1^{(i,j)},z_2^{(i,j)})\right|,\\
    \epsilon_\text{QAE}:=&\left|\frac{1}{N^2}\sum_{i,j=1}^{N-1}\theta(z_1^{(i,j)},z_2^{(i,j)})-\hat{\mu}\right|.
\end{align}
We start with the evaluation of $\epsilon_\text{trunc}$, which is the error associated with the truncation of the entire plane to the ring of inner radius $l$ and outer radius $L$. From the fact that $z_1^2+z_2^2 = -2\ln u$, we choose $l,L$ such that 
$ e^{-\frac{L^2}{2}}=u_\text{min}\equiv\frac{C_1}{2}\epsilon,\label{eq:Umin}$ and 
$
e^{-\frac{l^2}{2}}=u_\text{max}\equiv e^{-\frac{C_1}{2}\epsilon}\label{eq:Umax}$.
The outer and inner truncation error can be respectively bounded as follows
\begin{align}
    %\left |\int\int_{z_1^2+z_2^2\geq L^2}\theta(z_1,z_2)f_Z(z_1,z_2)dz_1dz_2\right| 
    \left | I - I(0,L) \right |
    &\leq \int\int_{z_1^2+z_2^2\geq L^2}f_Z(z_1,z_2)dz_1dz_2 = \int_0^{e^{\frac{-L^2}{2}}}\int_0^1 du dv = e^{\frac{-L^2}{2}}\;.\\
    %\left|\int\int_{z_1^2+z_2^2\leq l^2}\theta(z_1,z_2)f_Z(z_1,z_2)dz_1dz_2\right |
    \left | I(0,l) \right |
    &\leq \int\int_{z_1^2+z_2^2\leq l^2}\frac{1}{2\pi}dz_1dz_2 = \frac{l^2}{2}.
\end{align}
The choice in Eqs.~\eqref{eq:Umin} and \eqref{eq:Umax} obtains
\begin{align}
    \epsilon_\text{trunc}\le \frac{l^2}{2}+e^{-\frac{L^2}{2}}=C_1\epsilon\;.
\end{align}
The second error $\epsilon_\text{disc}$ is the discretization error and can be bounded by a Riemann sum bound. From the Box-Muller transform, we have 
\begin{align}
    &\left|I(l,L)-\frac{1}{N^2}\sum_{i,j=1}^{N-1}\theta(z_1^{(i,j)},z_2^{(i,j)})\right|=\left|\int_{u_\text{min}}^{u_\text{max}}\int_{v_\text{min}}^{v_\text{max}}\tilde{\theta}(u,v)dudv-\frac{1}{N^2}\sum_{i,j=1}^{N-1}\tilde{\theta}(u^{(i)},v^{(j)})\right|\label{eq:riemann},
\end{align}
where $\tilde{\theta}(u,v)=\theta(\sqrt{-2\ln u}\sin(2\pi v),\sqrt{-2\ln u}\cos(2\pi v))\equiv\theta(z_1,z_2)$, and we have used the property \eqref{eq:change of variables} for 
$\rho=0$ to change the integration variables.
If we define the bounds for the partial derivatives as
\begin{align}
    \frac{\partial^2 \tilde{\theta}}{\partial u^2}\le M_u,
    \frac{\partial^2 \tilde{\theta}}{\partial v^2}\le M_v,
\end{align}
we can apply Riemann sum in Appendix \ref{app:riemann2mid} and obtain the bound for Eq. \ref{eq:riemann} as
\begin{align}
    \left|\int_{u_\text{min}}^{u_\text{max}}\int_{v_\text{min}}^{v_\text{max}}\tilde{\theta}(u,v)dudv-\frac{1}{N^2}\sum_{i,j=1}^{N-1}\tilde{\theta}(u^{(i)},v^{(j)})\right|\le\frac{M_u+M_v}{24N^2}.
\end{align}
Define again $a:= \sin(2 \pi v)$ and $b:= \cos(2 \pi v)$. To find the upper bounds $M_u,M_v$, note that the first derivatives are as follows
\begin{align}
    &\frac{\partial \tilde{\theta}}{\partial u}
    =\frac{\partial {\theta}}{\partial z_1}\frac{\partial z_1}{\partial u}+\frac{\partial {\theta}}{\partial z_2}\frac{\partial z_2}{\partial u}
    =\frac{-1}{u\sqrt{-2\ln u}}\left(\frac{\partial {\theta}}{\partial z_1}a+\frac{\partial {\theta}}{\partial z_2}b\right),\\
    &\frac{\partial \tilde{\theta}}{\partial v}=\frac{\partial {\theta}}{\partial z_1}\frac{\partial z_1}{\partial v}+\frac{\partial {\theta}}{\partial z_2}\frac{\partial z_2}{\partial v}
    =2\pi\sqrt{-2\ln u}\left(\frac{\partial {\theta}}{\partial z_1}b-\frac{\partial {\theta}}{\partial z_2}a\right),
\end{align}
The second derivatives are as follows
\begin{align}
    \frac{\partial^2 \tilde{\theta}}{\partial u^2}
    &=\frac{1}{u^2\sqrt{-2\ln u}}\left(\frac{1}{-2\ln u}+1\right)\left(\frac{\partial {\theta}}{\partial z_1}a+\frac{\partial {\theta}}{\partial z_2}b \right)+\frac{1}{u^2(-2\ln u)}\left(\frac{\partial^2 \theta}{\partial z_1^2}a^2
    +\frac{\partial^2 \theta}{\partial z_1\partial z_2}ab
    +\frac{\partial^2 \theta}{\partial z_2^2}b^2\right)\nonumber\\
    &\le\frac{1}{u^2\sqrt{-2\ln u}}\left(\frac{1}{-2\ln u}+1\right)D_1+\frac{1}{u^2(-2\ln u)}D_2,
    \\
    \frac{\partial^2 \tilde{\theta}}{\partial v^2}
    &=-(2\pi)^2\sqrt{-2\ln u}\left(\frac{\partial {\theta}}{\partial z_1}a+\frac{\partial {\theta}}{\partial z_2}b\right)+(2\pi)^2(-2\ln u)\left(\frac{\partial^2 \theta}{\partial z_1^2}b^2
    -\frac{\partial^2 \theta}{\partial z_1\partial z_2}ab
    +\frac{\partial^2 \theta}{\partial z_2^2}a^2\right)\nonumber\\
    &\le-(2\pi)^2\sqrt{-2\ln u}D_1+(2\pi)^2(-2\ln u)D_2,
\end{align}
where $\left|\frac{\partial \theta}{\partial z_1}\right|+\left|\frac{\partial \theta}{\partial z_2}\right|\le D_1$ and $\left|\frac{\partial^2 \theta}{\partial z_1^2}\right|+\left|\frac{\partial^2 \theta}{\partial z_1\partial z_2}\right|+\left|\frac{\partial^2 \theta}{\partial z_2^2}\right|\le D_2$. The derivatives take the largest values at the edges of the domain of $u,v$. 
Expanding the $u,v$ dependent terms via a Taylor series and plugging in $u_\text{min}$ and $u_\text{max}$, we have
\begin{align}
    \nonumber \frac{\partial^2 \tilde{\theta}}{\partial u^2}\!\Biggm\vert_{u=u_\text{max}}
    =&\left(\frac{1}{\sqrt{2(1-u_\text{max})}}+O(1)\right)\left(\frac{1}{2\left(1-u_\text{max}\right)}+O(1)\right)D_1\\
    &+\left(\frac{1}{2\left(1-u_\text{max}\right)}+O(1)\right)D_2
    =\frac{D_1}{2\sqrt{2}}\left(\frac{\pi}{24}\epsilon\right)^{-\frac{3}{2}}+O\left(\epsilon^{-1}\right),\\
    \frac{\partial^2 \tilde{\theta}}{\partial u^2}\!\Biggm\vert_{u=u_\text{min}}
    =&\frac{1}{u_\text{min}^2\sqrt{-2\ln u_\text{min}}}\left(\frac{1}{{-2\ln u_\text{min}}}+1\right)D_1+\frac{1}{u_\text{min}^2(-2\ln u_\text{min})}D_2\nonumber\\
    =&\frac{D_1}{\left(\frac{1}{24}\epsilon\right)^2\sqrt{-2\ln \frac{\epsilon}{24}}}+O\left(\epsilon^{-2}\left(\ln\frac{1}{\epsilon}\right)^{-1}\right),\\
    \frac{\partial^2 \tilde{\theta}}{\partial v^2}\!\Biggm\vert_{u=u_\text{min}}=&-(2\pi)^2\sqrt{-2\ln u_\text{min}}D_1+(2\pi)^2(-2\ln u_\text{min})D_2\nonumber\\
    =&(2\pi)^2\left(-2\ln \frac{\epsilon}{24}\right)D_2+O\left(\left(\ln \frac{1}{\epsilon}\right)^\frac{1}{2}\right),
\end{align}
and therefore
\begin{align}
    &M_u=\frac{D_1}{\left(\frac{1}{24}\epsilon\right)^2\sqrt{-2\ln \frac{\epsilon}{24}}}+O\left(\epsilon^{-2}\left(\ln\frac{1}{\epsilon}\right)^{-1}\right),\\
    &M_v=(2\pi)^2\left(-2\ln \frac{\epsilon}{24}\right)D_2+O\left(\left(\ln \frac{1}{\epsilon}\right)^\frac{1}{2}\right).
\end{align}
If we focus on the leading terms, we obtain the discretization error as
\begin{align}
    \frac{M_u+M_v}{24N^2}\le\frac{C_2}{N^2\epsilon^2}=\epsilon_\text{disc},\label{eq:disc_error}
\end{align}
where $C_2$ is a constant.

The third QAE error can be bounded as 
\begin{align}
    \left|\frac{1}{N^2}\sum_{j,k=1}^{N-1}\theta(z_1^{(j,k)},z_2^{(j,k)})-\hat{\mu}\right|\le C_3\left(\frac{\sqrt{\mu}}{t}+\frac{1}{t^2}\right)=\epsilon_\text{QAE},\label{eq:qae_error}
\end{align}
where $C_3$ is a constant and $t$ is the number of query access, using Lemma \ref{lemma:case1}.

Combining these, we have
\begin{align}
    \left|\mathbb{E}[\theta(Z_1,Z_2)]-\tilde{\mu}\right|\le C_1\epsilon+\frac{C_2}{N^2\epsilon^2}+C_3\left(\frac{\sqrt{\mu}}{t}+\frac{1}{t^2}\right),
\end{align}
so if we take $C_1=1/3,N=\sqrt{3C_2}\epsilon^{-3/2}$ and $t=3C_3\sqrt{\mu}\epsilon^{-1}$, we can bound the total error with $\epsilon$.
\qed

\subsection{Case 2: $\theta\geq 0$ and $\E[\theta^2]\le B^2$}
In order to apply Algorithm \ref{alg:case1} for non-bounded $\theta$, we define 
\begin{equation}
    \theta_{x,y}=
    \begin{cases}
        \theta&\theta\in[x,y)\\
        0&\text{otherwise}
    \end{cases}
\end{equation}
to breakdown the range of $\theta$. Using this notation, the goal expectation value can be expressed as
\begin{align}
    \E[\theta]=\E[\theta_{0,1}]+\sum_{l=1}^\infty\E[\theta_{2^{l-1},2^l}].
\end{align}
Because $\E[\theta_{2^{l-1},2^l}/2^l]$ is bounded within $[0,1]$, we can apply Algorithm \ref{alg:case1} and obtain the following Algorithm \ref{alg:case2}.

\begin{algorithm}[h]
\caption{Estimate $\E[\theta(Z_1,Z_2)]$ for $\theta\geq 0$ and $\E[\theta^2]\le B^2$}\label{alg:case2}
\begin{algorithmic}[1]
\Require A function $\theta$ such that $\theta\geq 0$ and $\E[\theta^2]\le B^2$ and $\left|\frac{\partial \theta}{\partial z_1}\right|+\left|\frac{\partial \theta}{\partial z_2}\right|$ and $\left|\frac{\partial^2 \theta}{\partial z_1^2}\right|+\left|\frac{\partial^2 \theta}{\partial z_1\partial z_2}\right|+\left|\frac{\partial^2 \theta}{\partial z_2^2}\right|$ are bounded relatively to $\theta$, truncation points $U_\text{min},U_\text{max}$, $\epsilon, \delta>0$.
\begin{enumerate}
\item Set $k=\lceil\log_21/\tilde{\epsilon} \rceil,t_0=\lceil\tilde{D}\sqrt{\log_2 1/\tilde{\epsilon}}/\tilde{\epsilon}\rceil$ with $\tilde{\epsilon}=\frac{1}{3(B+1)^2}\epsilon$, $\epsilon_0=\epsilon$, and constants $D,\tilde{D}$.
\item Use Algorithm \ref{alg:case1} with $U_\text{min},U_\text{max},N=N_0=\left\lceil D\sqrt{k+1}/\epsilon^{3/2}\right\rceil,t=t_0$ with $\epsilon=\epsilon_0$ and $\delta = \delta/2$ to obtain estimate $\hat{\mu}_0$ for $\E[\theta_{0,1}]$.
\item For $l=1,\cdots,k$, use Algorithm \ref{alg:case1} with $U_\text{min},U_\text{max},N=N_l=\left\lceil D\sqrt{k+1}/\epsilon_l^{3/2}\right\rceil,t=t_0$ with $\epsilon=\epsilon_l=\frac{\epsilon}{2^l}$ and $\delta = \frac{\delta}{2k}$ to obtain estimate $\hat{\mu}_l$ for $\E[\theta_{2^{l-1},2^l}/2^l]$.
\item Output $\hat{\mu}=\hat{\mu_0}+\sum_{l=1}^k2^l\hat{\mu}_l$.
\end{enumerate}
\end{algorithmic}
\end{algorithm}

\begin{theorem}\label{thm:case2}
Let $\epsilon>0$ and assume a function $\theta$ such that $\E[\theta^2]\le B^2$ and $\left|\frac{\partial \theta}{\partial z_1}\right|+\left|\frac{\partial \theta}{\partial z_2}\right|$ and $\left|\frac{\partial^2 \theta}{\partial z_1^2}\right|+\left|\frac{\partial^2 \theta}{\partial z_1\partial z_2}\right|+\left|\frac{\partial^2 \theta}{\partial z_2^2}\right|$ are bounded with constant, relatively to $\theta$. Then Algorithm \ref{alg:case2} uses $N\in \tilde{O}(\epsilon^{-3/2}),t\in \tilde{O}(\epsilon^{-1})$ and outputs the estimate $\hat{\mu}$ such that
\begin{align}
    |\E[\theta(Z_1,Z_2)]-\hat{\mu}|\le \epsilon.\label{eq:error_algo2}
\end{align}
with probability at least $1-\delta$.
\end{theorem}
\proof The overall error can be bounded by the following inequality:
\begin{align}
    |\E[\theta]-\hat{\mu}|\le|\E[\theta_{0,1}]-\hat{\mu}_0|+\sum_{l=1}^k2^{l}|\E[\theta_{2^{l-1},2^l}/2^l]-\hat{\mu}_{l}|+\E[\theta_{2^k,\infty}].
\end{align}
Using Algorithm \ref{alg:case1}, we obtain the estimated $\hat{\mu}_0$ such that
\begin{align}
    |\E[\theta_{0,1}]-\hat{\mu}_0|\le C_1\epsilon_0+\frac{C_2}{N_0^2\epsilon_0^2}+C_3\left(\frac{\sqrt{\mu_0}}{t_0}+\frac{1}{t_0^2}\right).
\end{align}
Similarly, for $l=1,\cdots,k$, we can estimate $\hat{\mu}_l$ such that
\begin{align}
    |\E[\theta_{2^{l-1},2^l}/2^l]-\hat{\mu}_l|\le C_1\epsilon_l+\frac{C_2}{N_l^2\epsilon_l^2}+C_3\left(\frac{\sqrt{\mu_l}}{t_0}+\frac{1}{t_0^2}\right)
\end{align}
where $\mu_0 := \E[\theta_{0,1}]$ and $\mu_l := \E[\theta_{2^{l-1},2^l}/2^l]$.
Also we have
\begin{align}
    \E[\theta_{2^k,\infty}]\le\frac{\E[\theta^2]}{2^k}\le\frac{B^2}{2^k}.
\end{align}
Collecting these, we have
\begin{align}
    |\E[\theta]-\hat{\mu}|\le C_1\epsilon_0+\frac{C_2}{N_0^2\epsilon_0^2}+C_3(\frac{\sqrt{\mu_0}}{t_0}+\frac{1}{t_0^2})+\sum_{l=1}^k2^l\left(C_1\epsilon_l+\frac{C_2}{N_l^2\epsilon_l^2}+C_3\left(\frac{\sqrt{\mu_l}}{t_0}+\frac{1}{t_0^2}\right)\right)+\frac{B^2}{2^k}.\label{eq:bound_algo2}
\end{align}
\begin{comment}
The sixth term can be bounded by Cauchy-Schwarz as
\begin{align}
    \sum_{l=1}^k2^{l/2}\sqrt{\mu_l}
    \le\sqrt{k}\left(\sum_{l=1}^k2^l\mu_l\right)^{1/2}
    \le\sqrt{k}\left(\sum_{l=1}^k2^l\frac{\E[\theta^2_{2^{l-1},2^l}]}{2^{l-1}}\right)^{1/2}
    \le\sqrt{2k}\sigma.
\end{align}
\end{comment}
From Lemma \ref{lemma:case2}, the third, sixth, and seventh terms that correspond to QAE error can be bounded as follows, with probability at least $1-\delta$
\begin{align}
    C_3\left(\frac{\sqrt{\mu_0}}{t_0}+\frac{1}{t_0^2}\right)+\sum_{l=1}^k2^lC_3\left(\frac{\sqrt{\mu_l}}{t_0}+\frac{1}{t_0^2}\right)+\frac{B^2}{2^k}\le(B+1)^2\tilde{\epsilon}\label{eq:qae_algo2}
\end{align}
by taking $k=\lceil\log_21/\tilde{\epsilon} \rceil,t_0=\left\lceil\frac{D\sqrt{\log_2 1/\tilde{\epsilon}}}{\tilde{\epsilon}}\right\rceil$ for large enough constant $D$, so we set $\tilde{\epsilon}=\frac{1}{3(B+1)^2}\epsilon$ to bound Eq. \eqref{eq:qae_algo2} with $\epsilon/3$.
Also we take $C_1=\frac{1}{3(k+1)},u_\text{min}=C_1\epsilon_l/2,u_\text{max}=\exp(-C_1\pi\epsilon_l/2), N_0=N_l=\sqrt{3(k+1)C_2}\epsilon_l^{-3/2},\epsilon_l=\frac{1}{2^l}\epsilon$ and the remaining terms in Eq. \eqref{eq:bound_algo2} can be bounded as
\begin{align}
    C_1\epsilon_0+\frac{C_2}{N_0^2\epsilon_0^2}+\sum_{l=1}^k2^l\left(C_1\epsilon_l+\frac{C_2}{N_l^2\epsilon_l^2}\right)\le\frac{2}{3}\epsilon.
\end{align}
\qed

\subsection{Case 3: $\mathrm{Var}[\theta]\le \sigma^2$}

\begin{algorithm}[h]
\caption{Estimate $\E[\theta(X_1,X_2)]$ for $\mathrm{Var}[\theta]\le \sigma^2$}\label{alg:case3}
\begin{algorithmic}[1]
\Require $\mathrm{Var}[\theta]\le \sigma^2$.
\begin{enumerate}
\item Set $\theta'=\frac{\theta}{\sigma}$.
\item Obtain an output $\hat{\mu}'$ from $O_{\theta'}$
\item Let $\tilde{\theta}=\theta'-\hat{\mu}',\tilde{\theta}^-=\tilde{\theta}_{-\infty,0},\tilde{\theta}^+=\tilde{\theta}_{0,\infty}$.
\item Apply Algorithm \ref{alg:case2} for $-\tilde{\theta}^-/4$ and $\tilde{\theta}^+/4$ to obtain outputs $\hat{\mu}^-,\hat{\mu}^+$ with accuracy $\epsilon/(8\sigma)$ and failure probability 1/9
\item Output $\sigma\hat{\mu}=\sigma(\hat{\mu}'-4\hat{\mu}^-+4\hat{\mu}^+)$.
\end{enumerate}
\end{algorithmic}
\end{algorithm}

\begin{theorem}\label{thm:case3}
Let $0<\epsilon<4\sigma$ and assume a function $\theta$ such that $\mathrm{Var}[\theta]\le \sigma^2$ and $\left|\frac{\partial \theta}{\partial z_1}\right|+\left|\frac{\partial \theta}{\partial z_2}\right|$ and $\left|\frac{\partial^2 \theta}{\partial z_1^2}\right|+\left|\frac{\partial^2 \theta}{\partial z_1\partial z_2}\right|+\left|\frac{\partial^2 \theta}{\partial z_2^2}\right|$ are bounded with constant. Then Algorithm \ref{alg:case3} uses $N\in \tilde{O}((\epsilon/\sigma)^{-3/2}),t\in \tilde{O}((\epsilon/\sigma)^{-1})$ and outputs the estimate $\hat{\mu}$ such that
\begin{align}
    |\E[\theta(Z_1,Z_2)]-\hat{\mu}|\le \epsilon.\label{eq:error_algo3}
\end{align}
with success probability at least 2/3.
\end{theorem}
\proof 
Due to Chebyshev's inequality 
\begin{align}
    \mathrm{Pr}[|\theta'-{\mu}'|\ge3]\le\frac{1}{9},
\end{align}
so we assume $|\mu'-\hat{\mu}'|\le3$. Then we have, according to the triangle inequality and the fact that $\mathrm{Var}[\theta']\leq 1$
\begin{align}
    \E[\tilde{\theta}^2]^{1/2}=\E[(\theta'-\hat{\mu}')^2]^{1/2}\le\E[({\theta}'-\mu')^2]^{1/2}+\E[(\mu'-\hat{\mu}')^2]^{1/2}\le4,
\end{align}
and thus $\E[(-\tilde{\theta}^-/4)^2]\le1$ and $\E[(\tilde{\theta}^+/4)^2]\le1$.
Therefore, Algorithm \ref{alg:case2} can be applied to $-\tilde{\theta}^-/4$ and $\tilde{\theta}^+/4$ to obtain $\hat{\mu}^-,\hat{\mu}^+$ each with failure probability $1/9$ such that
\begin{align}
    \E[4\hat{\mu}^--\tilde{\theta}^-]\leq \frac{\epsilon}{2\sigma},\quad \E[4\hat{\mu}^+-\tilde{\theta}^+]\leq \frac{\epsilon}{2\sigma}.
\end{align}
By linearity of expectation, we have $\E[\sigma \hat{\mu}-\theta]\leq \epsilon$ with probability at least $2/3$.
\qed

\section{Resource estimate for quantum arithmetic operations}\label{section:arithmetic}
\subsection{Resources for arithmetic operations}
We use fixed-point arithmetic to represent real numbers 
\begin{align}
    x = \underbrace{x_{n-1}...x_{n-p}}_{p}.\underbrace{x_{n-p-1}...x_{0}}_{n-p}
\end{align}
and keep both $n$ and $p$ fixed over the course of computation.

 \textbf{Multiplication}: Following the method of \cite{Hner2018OptimizingQC}, the number of Toffoli gates required in a multiplication between two $(n,p)$ numbers is 
\begin{align}
    \text{T-count}_\text{mul}(n,p) = \frac{3}{2}n^2+3np+\frac{3}{2}n-3p^2+3p\;.\label{eq:mult}
\end{align}
The T-depth is given by \cite{Chakrabarti_2021}
\begin{align}
    \text{T-depth}_\text{mul}(n,p) = n\big(\text{T-depth}_\text{add}(n,p)+6\big)\;,
\end{align}
where 
\begin{align}
    \text{T-depth}_\text{add}(n,p) = \lfloor \log_2 (n) \rfloor + \lfloor \log_2 (n-1) \rfloor + \lfloor \log_2 (\frac{n}{3}) \rfloor + \lfloor \log_2 (\frac{n-1}{3}) \rfloor + 8\;
\end{align}
is the T-depth for addition.
%In this method the ancilla qubits can be reused. 
%A more recent work \cite{KahanamokuMeyer2024FastQI} makes use of no ancilla qubit.
% \item \textbf{Square root}: for the square root, we use the method of \cite{MuozCoreas2017TcountAQ}, which uses
% \begin{align}
%     \mathrm{Toffoli}_\text{sqr}(n,p) = \frac{n^2}{2}+3n-4\;.\label{eq:sqrt}
% \end{align}
% Toffoli gates and $2n+1$ qubits: $n$ for input, $n$ for output and $1$ for ancilla. 
% The good thing is the ancilla is restored to it's initial value after the computation. In this paper each Toffoli gate has a T-count of $7$ and a T-depth of $5$
% \begin{align}
%     \text{T}^{\text{count}}_\text{sqr}(n,p) &= \frac{7n^2}{2}+21n-28\;,\\
%     \text{T}^{\text{depth}}_\text{sqr}(n,p) &= 5n+3\;.
% \end{align}

 \textbf{Sine and Cosine}:
Using a piecewise polynomial approximation with $M$ pieces and degree $d$ as in \cite{Hner2018OptimizingQC}, the Toffoli count for implementing sine and cosine is given by
\begin{align}
    \text{T-count}(n,p,d,M)=\frac{3}{2}n^2d+3npd+\frac{7}{2}nd-3p^2d+3pd-d+2Md(4\lceil \log_2 M\rceil -8)+4Mn\;.\label{eq:pp}
\end{align} 
This method uses $n(d+1)+\lceil \log_2M\rceil +1$ qubits and has a T-depth of \cite{Chakrabarti_2021}
\begin{align}
    \text{T-depth}(n,p,d,M) = d\big(\text{T-depth}_\text{mul}(n,p)+\text{T-depth}_\text{add}(n,p)\big)+M\big(2\lfloor \log_2 (n-1) \rfloor+5\big)\;.
\end{align}
%Compare with the implementation of \cite{Cao_2013}.

 \textbf{Logarithm}: for the logarithm on the interval $[0,1]$ we cannot directly apply the polynomial approximation method of \cite{Hner2018OptimizingQC} as the derivative of the function diverges near $0$. One way to bypass this is to shift the argument by a certain power of two until it falls within the interval $[1/2,1]$
\begin{align}
\log x = \log x^* +k\log 2,
\end{align}
where $k := \lceil \log_2 x\rceil,x^* := x/2^k$. In \cite{10.5555/3179448.3179450} a simple circuit for finding $\lceil \log_2 x\rceil$ was shown, it involves no ancilla and uses $n$ CNOT gates. As a result, the dominant cost comes from two multiplications \eqref{eq:mult} and one logarithm, which is now well defined and given by \eqref{eq:pp}.

\textbf{Square root}: we proceed similarly as in the case of the logarithm. 

\subsection{Numerical experiment}
We now implement the above operations for small values of $n,p,d$ and $M$ and analyze the resulting error. Unlike other quantum state preparation methods, which prepare a distribution function in the amplitudes and thus the error can be given by some $L-$norm, the quantum Box-Muller transform prepares a set of samples $\lbrace Z_i\rbrace_{i=1}^N$ on the qubits instead, and we need an alternative metric to quantify the error arising from such a set. Here, we consider two error metrics
\begin{itemize}
    \item The relative error in estimating the expectation value of the exponential function
    \begin{align}
        \epsilon_\text{exp} := \frac{|\frac{1}{N}\sum \exp(Z_i)-\exp(1/2)|}{\exp(1/2)},\label{eq:exp_error}
    \end{align}
    where we have used the fact that $\E[\exp(Z)] = \exp(1/2)$ if $Z$ follows a standard normal distribution. The exponential function is crucial for financial applications such as derivative pricing since it results in a lognormal distribution \cite{Rebentrost_2018}. From a technical standpoint, the exponential function saturates the requirement in \cref{thm:case2}, namely its first and second derivative are bounded relatively to the function itself.
    \item The root mean square error of different quantiles
    \begin{align}
    \epsilon_\text{quantile} : = \sqrt{\sum_{p\in \lbrace 0.05, 0.1, ..., 0.95\rbrace} (q_\text{exact}(p)-q_\text{sample}(p))^2},\label{eq:quantile_error}
    \end{align}
    where $q_\text{exact}(p)$ is the $p-$quantile of the standard normal distribution while $q_\text{sample}$ is the $p-$quantile of the Box-Muller sample. We consider $19$ values of $p$, evenly spaced between $0.05$ and $0.95$. These quantiles are important in risk analysis as they correspond to Value at Risks of the samples \cite{Woerner_2019}.
    % The step function is also important in risk analysis as it can be used to compute the quantiles such as Value at Risks of the samples \cite{Woerner_2019}. 
% The exact expectation value is analytically computable for both functions, and we can compare our result with those analytical solutions.
\end{itemize}
Note that because we want to quantify the error in the preparation of $\ket{Z_i}$ (i.e. up to step 2 of Algorithm \ref{alg:case1}), the error and cost associated with the subsequent steps are not included in this analysis.
% We first choose parameters $d$ and $M$ such that the combination is the most cost-effective. We use the function $\cos(Z)$ for this parameterization for two reasons. First, the expectation value of this function can be exactly computed: $\mathbb{E}[\cos(Z)]=e^{-1/2}$ if $Z$ follows a standard normal distribution. Second, the cosine function remains bounded at infinity which minimizes the impact from the truncation error, allowing us to focus on the polynomial approximation error. Using the selected parameters $d$ and $M$, we compute the error in estimating the expectation value of several functions of practical relevance, in particular the exponential and step functions. 

\begin{figure}[H]
    \centering

    % --------- Group 1 ---------
    \begin{subfigure}[t]{0.48\textwidth}
        \centering
        \includegraphics[width=\linewidth,height=0.28\textheight,keepaspectratio]{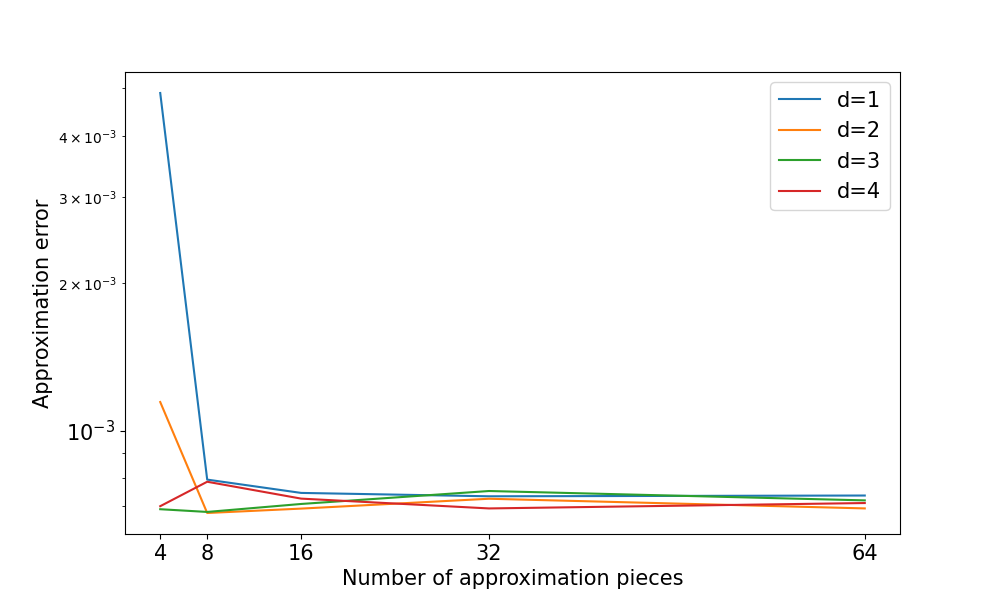}\caption*{$n=23$}
        \vspace{1mm}
        \includegraphics[width=\linewidth,height=0.28\textheight,keepaspectratio]{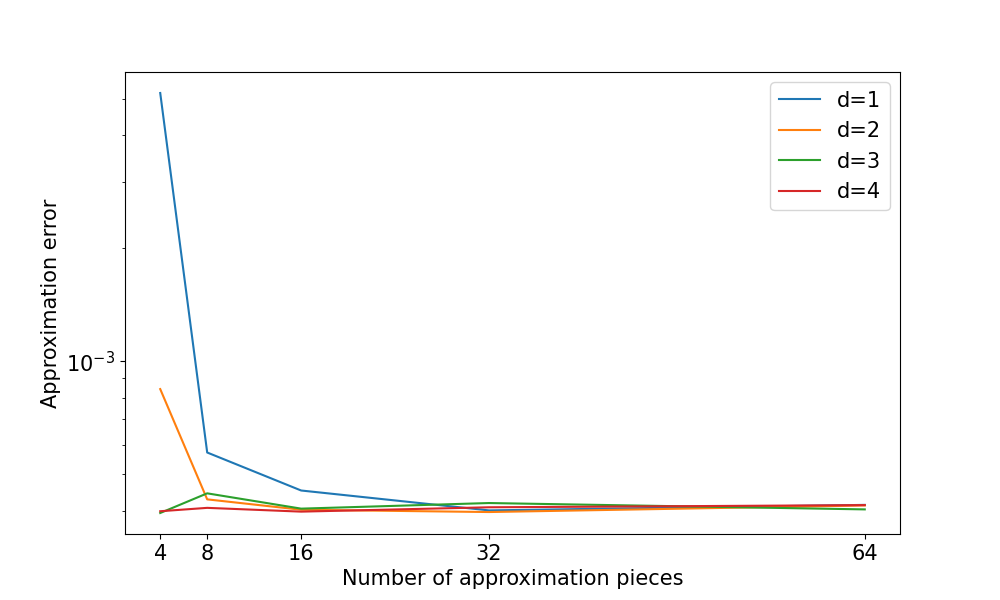}\caption*{$n=24$}
        \vspace{1mm}
        \includegraphics[width=\linewidth,height=0.28\textheight,keepaspectratio]{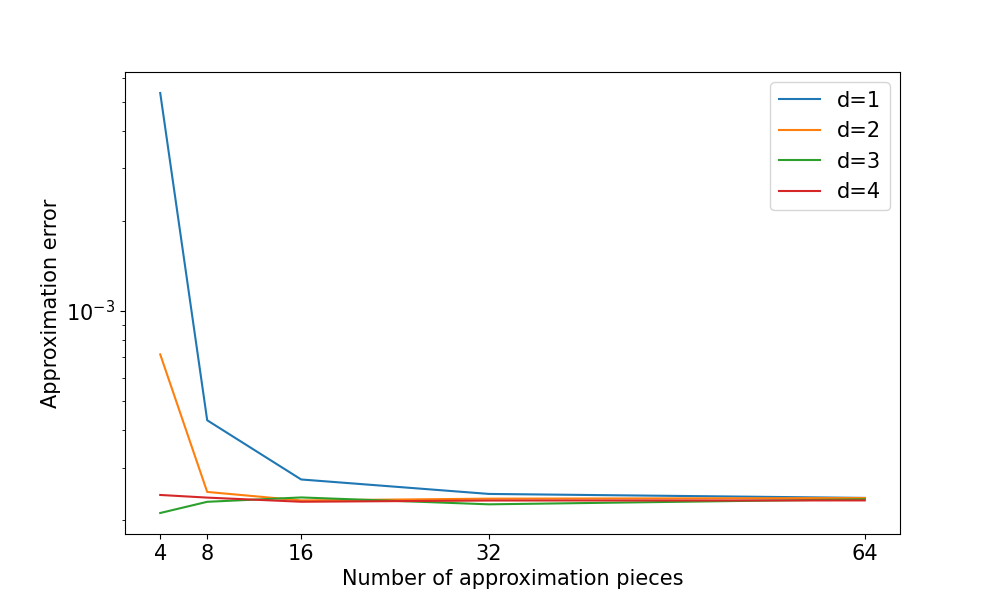}\caption*{$n=25$}
        \subcaption{Exponential error}
    \end{subfigure}
    \hfill
    % --------- Group 2 ---------
    \begin{subfigure}[t]{0.48\textwidth}
        \centering
        \includegraphics[width=\linewidth,height=0.28\textheight,keepaspectratio]{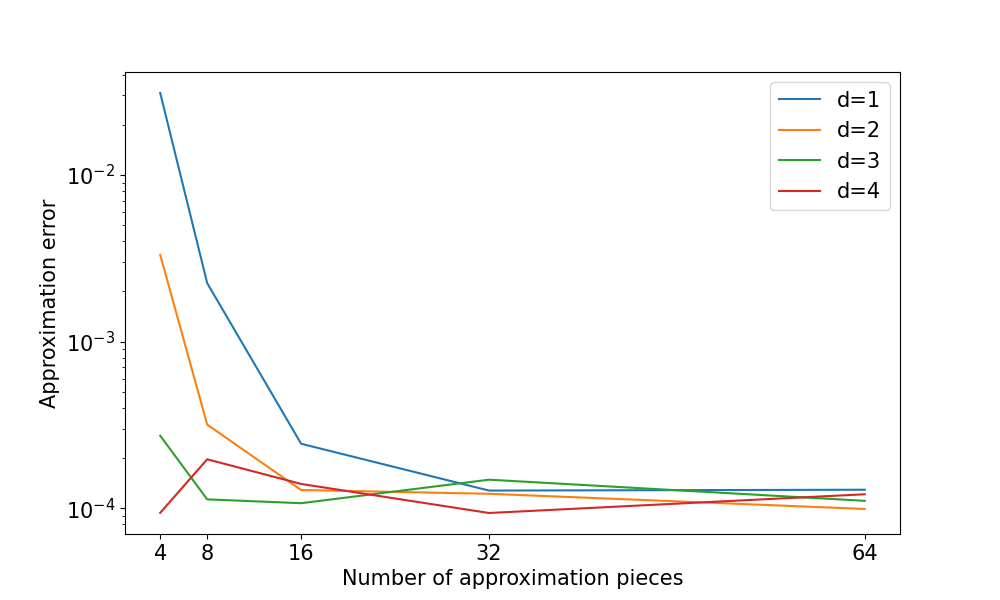}\caption*{$n=23$}
        \vspace{1mm}
        \includegraphics[width=\linewidth,height=0.28\textheight,keepaspectratio]{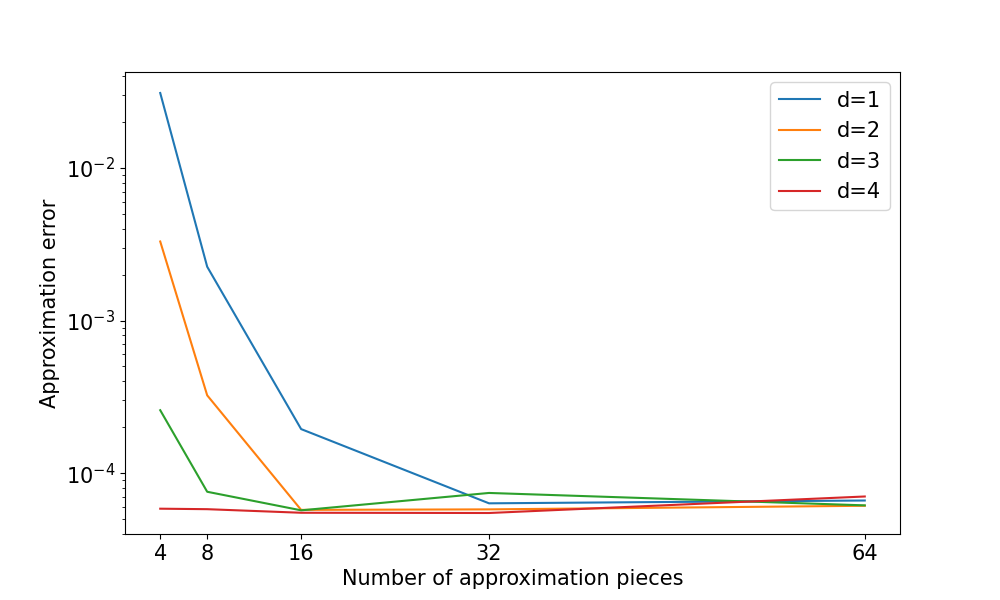}\caption*{$n=24$}
        \vspace{1mm}
        \includegraphics[width=\linewidth,height=0.28\textheight,keepaspectratio]{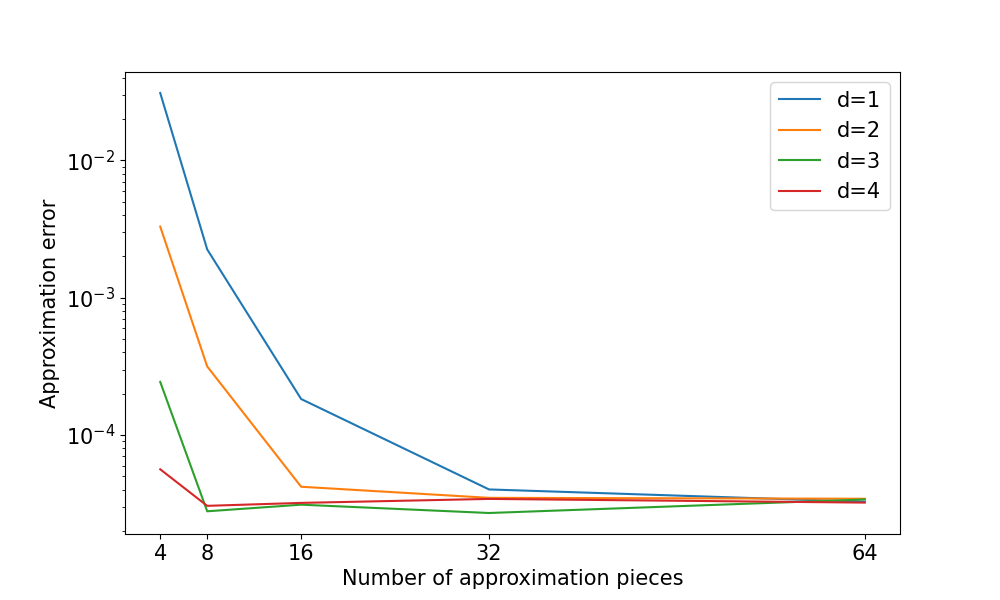}\caption*{$n=25$}
        \subcaption{Root mean square quantile error}
    \end{subfigure}

    \caption{The approximation error of the quantum Box-Muller transform for different parameter values. In this experiment, $p$ is set to $10$ to avoid overflow, $n\in \lbrace 20,21,22,23,24,25 \rbrace$, $d\in \lbrace 1,2,3,4\rbrace$, $M\in \lbrace 4,8,16,32,64\rbrace$. In all tested cases, the error saturates at $M=32$ and $d=1$. The improvement from higher degrees is insignificant as it is offset by the limited precision of the fixed-point representation: $2^{p-n}$.}
    \label{fig:error}
\end{figure}

The two error metrics are shown in \cref{fig:error} for various parameter values. In all cases, we set $p=10$ which is enough to avoid overflow. The number of pieces $M$ is set to be a power of two for several reasons. First of all, the number of qubits required for the polynomial approximation method depends on $\lceil \log_2 M\rceil$. Second, as can be directly seen from \cref{fig:error}, in most cases the error does not change significantly as we increase $M$ from $16$ to $64$. For this reason, it is rather redundant to consider more intermediate values. 

Since the number of qubits, the number of T gates and T depth all depend linearly in $d$, a small value of $d$ is preferred. Up to $n=25$, we found that the degree has no practical effect on the approximation error, as long as the number of pieces is sufficiently large. We thus set $M=32,d=1$ and present the errors, the total number of qubits, T-count and T-depth as functions of $n$ in \cref{table:resources}. For $d=1$, we found that it is sufficient to set $p=4$.
\begin{table}
\centering
\begin{tabular}{|c|c|c|c|c|c|c|}
\hline
    n&  Exp error & Quantile error &  N qubits & T-count & T-depth \\
    \hline
    10  & $2.797.10^{-2}$  &$1.408.10^{-2}$ & 78 & 8193 & 588 \\
    \hline
    11  & $1.634.10^{-2}$ & $4.735.10^{-3}$ &84 & 8942 & 610 \\
    \hline
    12  & $9.761.10^{-3}$ &$3.711.10^{-3}$ &90 & 9717 & 645 \\
    \hline
    13  & $6.042.10^{-3}$ & $1.438.10^{-3}$&96 & 10515 & 682 \\
    \hline
    14  &$3.897.10^{-3}$  & $9.814.10^{-4}$ & 102 & 11337 & 706 \\
    \hline
    15  & $2.122.10^{-3}$ &$4.876.10^{-4}$ &108 & 12183& 730\\
    \hline
    16  &$1.320.10^{-3}$ & $2.290.10^{-4}$ & 114 & 13053 & 771\\
    \hline
    17  & $9.913.10^{-4}$ & $1.327.10^{-4}$ & 120 & 13947 & 878\\
    \hline
    18  & $7.373.10^{-4}$ & $6.866.10^{-5}$&126 & 14865 & 904 \\
    \hline
    19  & $6.283.10^{-4}$  & $4.550.10^{-5}$&132 & 15807 & 930 \\
    \hline
\end{tabular}
\caption{Errors and quantum resources as functions of $n$, with $p=4, d=1$ and $M=32$. We stress that different errors are to be understood as different metrics to assert the approximation quality of a sample, and no construction of the exponential or indicator function is required.}
\label{table:resources}
\end{table}

\begin{figure}[h]
        \centering
        \begin{subfigure}[b]{0.45\textwidth}
            \centering
            \includegraphics[width=\textwidth]{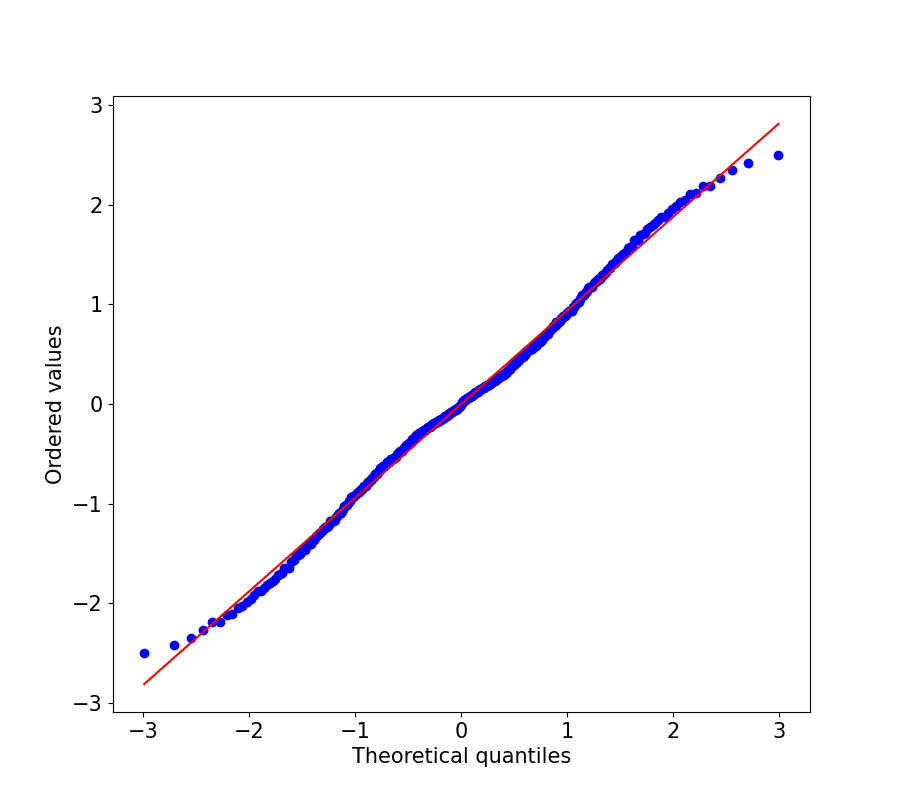}
            %\caption{The QQ plot} 
        \end{subfigure}
        \hfill
        \begin{subfigure}[b]{0.45\textwidth}  
            \centering 
            \includegraphics[width=\textwidth]{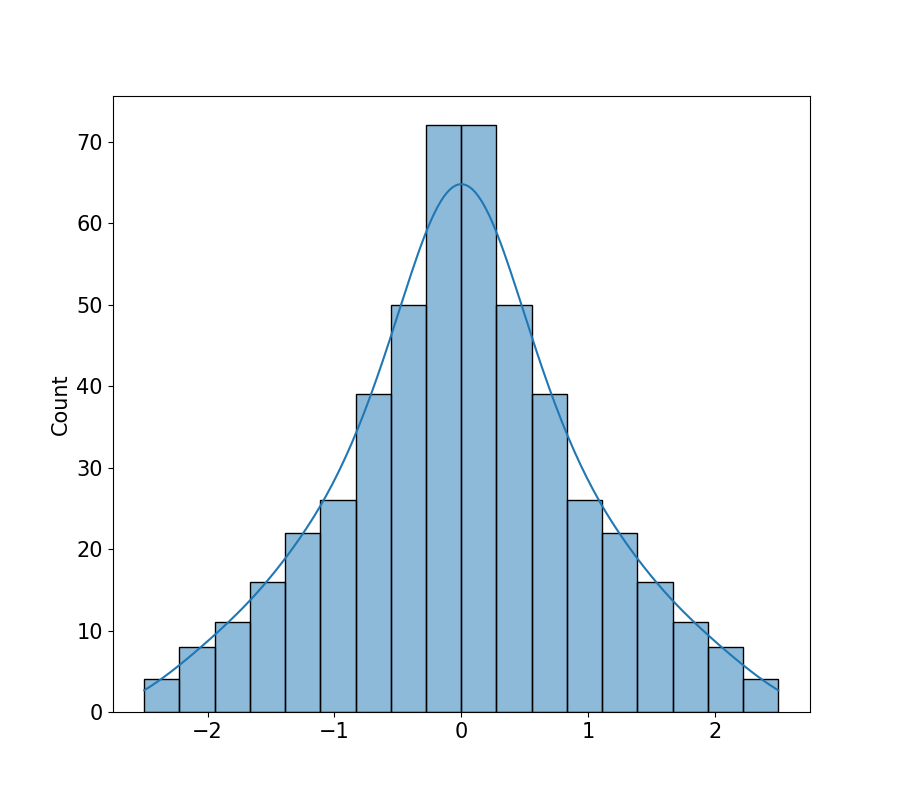}
            %\caption{The histogram} 
        \end{subfigure}
        \caption{The QQ plot and the histogram of 1024 samples from a Qiskit simulation of the quantum Box-Muller transform. This small experiment uses $5$ qubits to represent each uniform sample.}
        \label{fig:samples}
\end{figure}

We also simulate a simplified example using Qiskit. To minimize the resources, we adopt the following approximations
\begin{align}
    \sin(2\pi x)\approx \frac{1}{8}+4x\quad \text{for}\quad x\in [0,\frac{1}{4}]\quad ;\quad \sqrt{-2\ln x}\approx 2.5(1-x)\quad \text{for}\quad x\in [0,1]
\end{align}
The $\sin$ function is then extended to $x\in [0,1]$ by symmetry. In another word, we set $M$ and $d$ both to $1$, their smallest possible value and we choose approximating polynomials with simple coefficients. Since Qiskit does not support fixed-point representation, we have to work with integer arithmetics and then rescale everything back to real numbers. We use 5 qubits to represent each uniform sample ($32$ integers from $0$ to $31$ are rescaled to $32$ equidistant numbers from $0$ to $31/32$). The total number of qubits used in this experiment is $33$. The quality of the resulting 1024 Box-Muller samples is shown in \cref{fig:samples}

\section{Discussion}
Our algorithm gives a simple alternative to the standard state preparation-based approach, removing the black box assumption for state preparation.
However, compared to the other approach, our algorithm requires a slightly larger number of qubits $n$. To this end, let us take a closer look at the error analysis  in \cite{Chakrabarti_2021} and the one in this work.
% This can be seen from the discretization error in Eq. \eqref{eq:disc_error}, where we have $\epsilon_{\mathrm{disc}}\in O(1/(N^2\epsilon^2))$.
% When we assume $\epsilon_{\mathrm{disc}}\in O(\epsilon)$, this results in $N\in O(1/\epsilon^{3/2})$.
% In the state preparation approach, the discretization error is expressed as $\epsilon_{\mathrm{disc}}\in O(1/N^2)$, and thus $N\in O(1/\epsilon^{1/2})$ \cite[Eq. (23)]{Chakrabarti_2021}.

In both papers, the discretization error is obtained using similar technique and appears similar in the first glance.  In our paper, we have obtained
\begin{align}
\epsilon_\text{disc}\leq \frac{M_u+M_v}{24N^2},
\end{align}
where $M_u$ and $M_v$ are upper bound on the second derivative of the integrand with respect to the uniform variable $u$ and $v$. Due to the bound imposed on the truncation error, $M_u+M_v\in O(1/\epsilon^2)$, and thus we have the overall dependence $N\in O(1/\epsilon^{3/2})$. In \cite{Chakrabarti_2021}, it was obtained that
\begin{align}
    \epsilon_\text{disc}\leq \frac{\beta(2\omega \sigma_\text{max})^{dT+2}}{24N^2},
\end{align}
where $\beta$ is an upper bound on the second derivatives of the integrand with respect to the Gaussian variable, $2\omega \sigma_\text{max}$ is the truncated interval, and $dT$ is the number of dimensions.  For most financial applications, the integrand is a function of the log-normal distribution and its second derivative is exponentially large in $\omega\sigma_\text{max}$. The dependence of $\omega$ on $\epsilon$ follows from the Chernoff bounds and is given by $\epsilon_\text{trunc}\leq 2dT e^{-\omega^2/2}$. A rough estimate for an upper bound $\beta \in O(1/\epsilon)$ leads to a slightly more favorable result $N\in \tilde{O}(1/\epsilon)$. Note that the number of qubits needed for a given discretization error only scales as the logarithm of $N$. Hence, the difference of our method and the standard method in terms of the number of qubits needed for achieving a certain discretization error is less significant.

%Qualitative discussion/comparison to standard method with Gaussian state preparation assumption. We are slightly worse. However, we remove the black box assumption for state preparation. 

\section{Conclusion}\label{sec:conclusion}
To our knowledge, this work is the first to discuss the application of the Box-Muller transformation in a quantum algorithm as an alternative to state preparation. It is also the first method that only uses a constant number of quantum arithmetic operations to prepare a normal distribution, leading to a small gate complexity.
Because of the simplicity of Box-Muller transform, an upper bound of the integration error can be analytically obtained.
As Gaussian sampling is ubiquitous, our method may find application in many tasks as a subroutine.
\section*{Acknowledgments}
The authors acknowledge discussions with Johann Fong Cheok Arn and Darien Oh.
This project is supported by the National Research Foundation, Singapore through the National Quantum Office, hosted in A*STAR, under its Centre for Quantum Technologies Funding Initiative (S24Q2d0009).
CQT acknowledges funding from OCBC via a joint NUS-OCBC research project.
\bibliographystyle{utphys}
\bibliography{bm}

@article{Rebentrost_2018,
   title={Quantum computational finance: Monte Carlo pricing of financial derivatives},
   volume={98},
   ISSN={2469-9934},
   url={http://dx.doi.org/10.1103/PhysRevA.98.022321},
   DOI={10.1103/physreva.98.022321},
   number={2},
   journal={Physical Review A},
   publisher={American Physical Society (APS)},
   author={Rebentrost, Patrick and Gupt, Brajesh and Bromley, Thomas R.},
   year={2018},
   month=aug }

@misc{kitaev2009,
      title={Wavefunction preparation and resampling using a quantum computer}, 
      author={Alexei Kitaev and William A. Webb},
      year={2009},
      eprint={0801.0342},
      archivePrefix={arXiv},
      primaryClass={quant-ph},
      url={https://arxiv.org/abs/0801.0342}, 
}

@misc{mcardle2022quantumstatepreparationcoherent,
      title={Quantum state preparation without coherent arithmetic}, 
      author={Sam McArdle and András Gilyén and Mario Berta},
      year={2022},
      eprint={2210.14892},
      archivePrefix={arXiv},
      primaryClass={quant-ph},
      url={https://arxiv.org/abs/2210.14892}, 
}

@misc{mori2024efficientstatepreparationmultivariate,
      title={Efficient state preparation for multivariate Monte Carlo simulation}, 
      author={Hitomi Mori and Kosuke Mitarai and Keisuke Fujii},
      year={2024},
      eprint={2409.07336},
      archivePrefix={arXiv},
      primaryClass={quant-ph},
      url={https://arxiv.org/abs/2409.07336}, 
}

@misc{grover2002creating,
      title={Creating superpositions that correspond to efficiently integrable probability distributions}, 
      author={Lov Grover and Terry Rudolph},
      year={2002},
      eprint={quant-ph/0208112},
      archivePrefix={arXiv},
      primaryClass={quant-ph}
}

@misc{rosenkranz2024quantum,
      title={Quantum state preparation for multivariate functions}, 
      author={Matthias Rosenkranz and Eric Brunner and Gabriel Marin-Sanchez and Nathan Fitzpatrick and Silas Dilkes and Yao Tang and Yuta Kikuchi and Marcello Benedetti},
      year={2024},
      eprint={2405.21058},
      archivePrefix={arXiv},
      primaryClass={quant-ph}
}

@article{Moosa_2023,
   title={Linear-depth quantum circuits for loading Fourier approximations of arbitrary functions},
   volume={9},
   ISSN={2058-9565},
   url={http://dx.doi.org/10.1088/2058-9565/acfc62},
   DOI={10.1088/2058-9565/acfc62},
   number={1},
   journal={Quantum Science and Technology},
   publisher={IOP Publishing},
   author={Moosa, Mudassir and Watts, Thomas W and Chen, Yiyou and Sarma, Abhijat and McMahon, Peter L},
   year={2023}, pages={015002} }

@misc{rattew2022preparing,
      title={Preparing Arbitrary Continuous Functions in Quantum Registers With Logarithmic Complexity}, 
      author={Arthur G. Rattew and Bálint Koczor},
      year={2022},
      eprint={2205.00519},
      archivePrefix={arXiv},
      primaryClass={quant-ph}
}

@article{Grover_2000,
  title = {Synthesis of Quantum Superpositions by Quantum Computation},
  author = {Grover, Lov K.},
  journal = {Phys. Rev. Lett.},
  volume = {85},
  issue = {6},
  pages = {1334--1337},
  numpages = {0},
  year = {2000},
  publisher = {American Physical Society},
  doi = {10.1103/PhysRevLett.85.1334},
  url = {https://link.aps.org/doi/10.1103/PhysRevLett.85.1334}
}

@article{Montanaro_2015,
   title={Quantum speedup of Monte Carlo methods},
   volume={471},
   ISSN={1471-2946},
   url={http://dx.doi.org/10.1098/rspa.2015.0301},
   DOI={10.1098/rspa.2015.0301},
   number={2181},
   journal={Proceedings of the Royal Society A: Mathematical, Physical and Engineering Sciences},
   publisher={The Royal Society},
   author={Montanaro, Ashley},
   year={2015},
   month=sep, pages={20150301} }

@article{Hner2018OptimizingQC,
  title={Optimizing Quantum Circuits for Arithmetic},
  author={Thomas H{\"a}ner and Martin R{\"o}tteler and Krysta Marie Svore},
  journal={ArXiv},
  year={2018},
  volume={abs/1805.12445},
  url={https://api.semanticscholar.org/CorpusID:44124540}
}

@article{Black1973,
  title = {The pricing of options and corporate liabilities},
  author = {Black, Fischer and Scholes, Myron},
 journal = {Journal of Political Economy},
  volume = {81},
  issue = {3},
 pages = {637-654},
  year = {1973}
}

@article{Merton1973,
  author = {Robert C. Merton},
 journal = {The Bell Journal of Economics and Management Science},
 number = {1},
 pages = {141--183},
 title = {Theory of rational option pricing},
 volume = {4},
 year = {1973}
}

@article{Zoufal_2019,
   title={Quantum Generative Adversarial Networks for learning and loading random distributions},
   volume={5},
   ISSN={2056-6387},
   url={http://dx.doi.org/10.1038/s41534-019-0223-2},
   DOI={10.1038/s41534-019-0223-2},
   number={1},
   journal={npj Quantum Information},
   publisher={Springer Science and Business Media LLC},
   author={Zoufal, Christa and Lucchi, Aurélien and Woerner, Stefan},
   year={2019},
   month=nov }

@article{Chakrabarti_2021,
   title={A Threshold for Quantum Advantage in Derivative Pricing},
   volume={5},
   ISSN={2521-327X},
   url={http://dx.doi.org/10.22331/q-2021-06-01-463},
   DOI={10.22331/q-2021-06-01-463},
   journal={Quantum},
   publisher={Verein zur Forderung des Open Access Publizierens in den Quantenwissenschaften},
   author={Chakrabarti, Shouvanik and Krishnakumar, Rajiv and Mazzola, Guglielmo and Stamatopoulos, Nikitas and Woerner, Stefan and Zeng, William J.},
   year={2021},
   month=jun, pages={463} }

@article{10.5555/3179448.3179450,
author = {Bhaskar, Mihir K. and Hadfield, Stuart and Papageorgiou, Anargyros and Petras, Iasonas},
title = {Quantum algorithms and circuits for scientific computing},
year = {2016},
issue_date = {March 2016},
publisher = {Rinton Press, Incorporated},
address = {Paramus, NJ},
volume = {16},
number = {3–4},
issn = {1533-7146},
journal = {Quantum Info. Comput.},
month = mar,
pages = {197–236},
numpages = {40}
}

@article{Stamatopoulos_2020,
   title={Option Pricing using Quantum Computers},
   volume={4},
   ISSN={2521-327X},
   url={http://dx.doi.org/10.22331/q-2020-07-06-291},
   DOI={10.22331/q-2020-07-06-291},
   journal={Quantum},
   publisher={Verein zur Forderung des Open Access Publizierens in den Quantenwissenschaften},
   author={Stamatopoulos, Nikitas and Egger, Daniel J. and Sun, Yue and Zoufal, Christa and Iten, Raban and Shen, Ning and Woerner, Stefan},
   year={2020},
   month=jul, pages={291} }

@article{Herbert_2022,
   title={Quantum Monte Carlo Integration: The Full Advantage in Minimal Circuit Depth},
   volume={6},
   ISSN={2521-327X},
   url={http://dx.doi.org/10.22331/q-2022-09-29-823},
   DOI={10.22331/q-2022-09-29-823},
   journal={Quantum},
   publisher={Verein zur Forderung des Open Access Publizierens in den Quantenwissenschaften},
   author={Herbert, Steven},
   year={2022},
   month=sep, pages={823} }

@article{Woerner_2019,
   title={Quantum risk analysis},
   volume={5},
   ISSN={2056-6387},
   url={http://dx.doi.org/10.1038/s41534-019-0130-6},
   DOI={10.1038/s41534-019-0130-6},
   number={1},
   journal={npj Quantum Information},
   publisher={Springer Science and Business Media LLC},
   author={Woerner, Stefan and Egger, Daniel J.},
   year={2019},
   month=feb }

\newpage
\appendix

\section{Riemann sum}

\subsubsection{1D Right Riemann sum}
Let $x_k=a+\Delta k$ for $k=1,\cdots,n$ and $x_0=a,x_n=b$. Suppose that we have $|f'(x)|\le M_1$ for $x\in[a,b]$.
\begin{align}
    \left|\int_a^bf(x)dx-\sum_{i=1}^nf(x_{i})\Delta\right|\le M_1\frac{(b-a)^2}{2n}
\end{align}
\proof
\begin{align*}
    \left|\int_a^bf(x)dx-\sum_{i=1}^nf(x_{i})\Delta\right|
    &=\left|\sum_{i=1}^n\int_{x_{i-1}}^{x_i}f(x)dx-\sum_{i=1}^nf(x_{i})\Delta\right|\\
    &=\left|\sum_{i=1}^n\int_{x_{i-1}}^{x_i}f(x)dx-\sum_{i=1}^n\int_{x_{i-1}}^{x_i}f(x_i)dx\right|\\
    &=\left|\sum_{i=1}^n\int_{x_{i-1}}^{x_i}(f(x)-f(x_i))dx\right|\\
    &=\left|\sum_{i=1}^n\int_{x_{i-1}}^{x_i}f'(c)(x_i-x)dx\right|,
\end{align*}
where the last equality is from the mean value theorem and $c\in[x_{i-1},x_i]$. Now this amount can be bounded as 
\begin{align}
    \left|\sum_{i=1}^n\int_{x_{i-1}}^{x_i}f'(c)(x_i-x)dx\right|
    &\le\sum_{i=1}^n\int_{x_{i-1}}^{x_i}|f'(c)|(x_i-x)dx\\
    &\le M_1\sum_{i=1}^n\int_{x_{i-1}}^{x_i}(x_i-x)dx=M_1\sum_{i=1}^n\frac{\Delta^2}{2}=M_1\frac{(b-a)^2}{2n}.
\end{align}
\qed

\subsubsection{2D Right Riemann sum}\label{app:riemann2right}
Let $x_k=a+\Delta k$ for $k=1,\cdots,n$ and $x_0=a,x_n=b$. Same for $y_k$. Suppose that we have $|\frac{\partial}{\partial x}f(x,y)|\le M_x,|\frac{\partial}{\partial y}f(x,y)|\le M_y$ for $x,y\in[a,b]$.
\begin{align}
    \left|\int_a^b\int_a^bf(x,y)dxdy-\sum_{i,j=1}^nf(x_{i},y_j)\Delta^2\right|\le(M_x+M_y)\frac{(b-a)^3}{2n}
\end{align}
\proof
\begin{align*}
    \left|\int_a^b\int_a^bf(x,y)dxdy-\sum_{i,j=1}^nf(x_{i},y_j)\Delta^2\right|
    &=\left|\sum_{i,j=1}^n\int_{x_{i-1}}^{x_i}\int_{y_{j-1}}^{y_j}f(x,y)dxdy-\sum_{i,j=1}^nf(x_{i},y_j)\Delta^2\right|\\
    &=\left|\sum_{i,j=1}^n\int_{x_{i-1}}^{x_i}\int_{y_{j-1}}^{y_j}f(x,y)dxdy-\sum_{i,j=1}^n\int_{x_{i-1}}^{x_i}\int_{y_{j-1}}^{y_j}f(x_i,y_j)dxdy\right|\\
    &=\left|\sum_{i,j=1}^n\int_{x_{i-1}}^{x_i}\int_{y_{j-1}}^{y_j}(f(x,y)-f(x_i,y_j))dxdy\right|\\
    &=\left|\sum_{i,j=1}^n\int_{x_{i-1}}^{x_i}\int_{y_{j-1}}^{y_j}\left(\frac{\partial f}{\partial x}(c,d)(x_i-x)+\frac{\partial f}{\partial y}(c,d)(y_i-y)\right)dxdy\right|,
\end{align*}
where the last equality is from the mean value theorem and $c\in[x_{i-1},x_i],d\in[y_{j-1},y_j]$. Now this amount can be bounded as 
\begin{align}
    &\left|\sum_{i,j=1}^n\int_{x_{i-1}}^{x_i}\int_{y_{j-1}}^{y_j}\left(\frac{\partial f}{\partial x}(c,d)(x_i-x)+\frac{\partial f}{\partial y}(c,d)(y_i-y)\right)dxdy\right|\\
    &\le \sum_{i,j=1}^n\left(M_x\int_{x_{i-1}}^{x_i}\int_{y_{j-1}}^{y_j}(x_i-x)dxdy+M_y\int_{x_{i-1}}^{x_i}\int_{y_{j-1}}^{y_j}(y_j-y)dxdy\right)=(M_x+M_y)\frac{(b-a)^3}{2n}.
\end{align}
\qed

\subsubsection{1D Mid-point Riemann sum}
Let $\bar{x}_k=a+(k-\frac{1}{2})\Delta$ for $k=1,\cdots,n$ and $\Delta=(b-a)/n$. Suppose that we have $|f^{(2)}(x)|\le M_2$ for $x\in[a,b]$.
\begin{align}
    \left|\int_a^bf(x)dx-\sum_{i=1}^nf(\bar{x}_i)\Delta\right|\le M_2\frac{(b-a)^3}{24n^2}+O\left(\frac{1}{n^4}\right)
\end{align}
\proof
\begin{align*}
    \left|\int_a^bf(x)dx-\sum_{i=1}^nf(\bar{x}_i)\Delta\right|
    &=\left|\sum_{i=1}^n\int_{x_{i-1}}^{x_i}f(x)dx-\sum_{i=1}^n\int_{x_{i-1}}^{x_i}f(\bar{x}_i)dx\right|\\
    &=\left|\sum_{i=1}^n\int_{x_{i-1}}^{x_i}(f(x)-f(\bar{x}_i))dx\right|\\
    &=\left|\sum_{i=1}^n\int_{x_{i-1}}^{x_i}\left(f'(\bar{x}_i)(x-\bar{x}_i)+\frac{f^{(2)}(\bar{x}_i)}{2}(x-\bar{x}_i)^2+\frac{f^{(3)}(\bar{x}_i)}{3!}(x-\bar{x}_i)^3+O(x^4)\right)dx\right|\\
    &=\left|\frac{f^{(2)}(\bar{x}_i)(b-a)^3}{24n^2}+O\left(\frac{1}{n^4}\right)\right|,
\end{align*}
where in the third equality we used Taylor expansion around $\bar{x}_i$ and in the last equality we used that odd terms vanish because of symmetry.
\qed

\subsubsection{2D Mid-point Riemann sum}\label{app:riemann2mid}
Let $\bar{x}_k=a+(k-\frac{1}{2})\Delta$ for $k=1,\cdots,n$ and $\Delta=(b-a)/n$. Same for $\bar{y}_k$.
Suppose that we have $|\frac{\partial^2}{\partial x^2}f(x,y)|\le M_x,|\frac{\partial^2}{\partial y^2}f(x,y)|\le M_y$ for $x,y\in[a,b]$.
\begin{align}
    \left|\int_a^b\int_a^bf(x,y)dxdy-\sum_{i,j=1}^nf(\bar{x}_{i},\bar{y}_j)\Delta^2\right|\le(M_x+M_y)\frac{(b-a)^4}{24n^2}+O\left(\frac{1}{n^4}\right)
\end{align}
\proof
\begin{align*}
    &\left|\int_a^b\int_a^bf(x,y)dxdy-\sum_{i,j=1}^nf(\bar{x}_{i},\bar{y}_j)\Delta^2\right|\\
    &=\left|\sum_{i,j=1}^n\int_{x_{i-1}}^{x_i}\int_{y_{j-1}}^{y_j}f(x,y)dxdy-\sum_{i,j=1}^n\int_{x_{i-1}}^{x_i}\int_{y_{j-1}}^{y_j}f(\bar{x}_i,\bar{y}_j)dxdy\right|\\
    &=\left|\sum_{i,j=1}^n\int_{x_{i-1}}^{x_i}\int_{y_{j-1}}^{y_j}(f(x,y)-f(\bar{x}_i,\bar{y}_j))dxdy\right|\\
    &=\left|\sum_{i,j=1}^n\int_{x_{i-1}}^{x_i}\int_{y_{j-1}}^{y_j}\left(\frac{1}{2}\left(\frac{\partial^2f}{\partial x^2}(\bar{x}_i,\bar{y}_j)(x-\bar{x}_i)^2
    +\frac{\partial^2f}{\partial y^2}(\bar{x}_i,\bar{y}_j)(y-\bar{y}_j)^2\right)+\cdots\right)dxdy\right|\\
    &=\left|\left(\frac{\partial^2f}{\partial x^2}(\bar{x}_i,\bar{y}_j)+\frac{\partial^2f}{\partial y^2}(\bar{x}_i,\bar{y}_j)\right)\frac{(b-a)^4}{24n^2}+O\left(\frac{1}{n^4}\right)\right|,
\end{align*}
where in the third equality we used Taylor expansion around $\bar{x}_i$ but skipping the odd terms as they will be integrated out because of symmetry.
\qed

\end{document}